\documentclass[twocolumn,twocolappendix]{aastex631}

\shorttitle{VVV-WIT-12 and  its changing nebula: What  is this?}
\shortauthors{Saito et al.}

\graphicspath{{./}{figures/}}

\begin{document}

\title{VVV-WIT-12 and its fashionable nebula:  a four year long period
  \\ Young Stellar Object with a light echo?}

\author[0000-0001-6878-8648]{Roberto             K.             Saito}
\affiliation{Departamento  de Física,  Universidade  Federal de  Santa
  Catarina, Trindade 88040-900, Florianópolis, Brazil}

\author{Bringfried Stecklum}
\affiliation{Th\"uringer   Landessternwarte,    Sternwarte   5,   07778
  Tautenburg, Germany}

\author{Dante   Minniti}   
\affiliation{Instituto de Astrof\'isica, Dep. de Ciencias F{\'i}sicas,
  Facultad  de   Ciencias  Exactas,  Universidad  Andres   Bello,  Av.
  Fern\'andez Concha 700, Santiago, Chile}
\affiliation{Vatican Observatory, V00120 Vatican City State, Italy}
\affiliation{Departamento  de Física,  Universidade  Federal de  Santa
  Catarina, Trindade 88040-900, Florianópolis, Brazil}

\author{Philip W. Lucas}
\affiliation{Centre  for  Astrophysics, University  of  Hertfordshire,
  College Lane, Hatfield AL10 9AB, UK}

\author{Zhen Guo}
\affiliation{Instituto de F{\'i}sica  y Astronom{\'i}a, Universidad de
  Valpara{\'i}so,   ave.  Gran   Breta{\~n}a,   1111,  Casilla   5030,
  Valpara{\'i}so, Chile}
\affiliation{N\'ucleo   Milenio  de   Formaci\'on  Planetaria   (NPF),
  ave. Gran Breta{\~n}a, 1111, Casilla 5030, Valpara{\'i}so, Chile}
\affiliation{Departamento   de  F{\'i}sica,   Universidad  Tecnic{\'a}
  Federico Santa  Mar{\'i}a, Avenida Espa{\~n}a  1680, Valpara{\'i}so,
  Chile}

\author{Leigh C. Smith}
\affiliation{Institute   of   Astronomy,  University   of   Cambridge,
  Madingley Road, Cambridge CB3 0HA, UK}

\author{Luciano Fraga}
\affiliation{Laboratorio Nacional de Astrof\'isica LNA/MCTI, 37504-364
  Itajub\'a, MG, Brazil}

\author{Felipe Navarete}
\affiliation{SOAR Telescope/NSF's  NOIRLab, Avda Juan  Cisternas 1500,
  1700000, La Serena, Chile}

\author{Juan Carlos Beam\'in} 
\affiliation{Instituto   de    Astrof\'isica,   Dep.     de   Ciencias
  F{\'i}sicas, Facultad de Ciencias Exactas, Universidad Andres Bello,
  Av.  Fern\'andez Concha 700, Santiago, Chile}

\author{Calum Morris}
\affiliation{Centre  for  Astrophysics, University  of  Hertfordshire,
  College Lane, Hatfield AL10 9AB, UK}

\begin{abstract}

We  report  the  serendipitous  discovery of  VVV-WIT-12,  an  unusual
variable source  that seems to  induce variability in  its surrounding
nebula.  The  source belongs  to the  rare objects  that we  call WITs
(short for What Is This?) discovered within the VISTA Variables in the
V\'ia L\'actea (VVV) survey. VVV-WIT-12  was discovered during a pilot
search for light echoes from distant Supernovae (SNe) in the Milky Way
using  the near-IR  images  of the  VVV survey.   This  source has  an
extremely  red spectral  energy distribution,  consistent with  a very
reddened ($A_V \sim  100$ mag) long period  variable star ($P\sim1525$
days).   Furthermore,  it  is  enshrouded in  a  nebula  that  changes
brightness and color  with time, apparently in synch  with the central
source  variations.    The  near-IR  light  curve   and  complementary
follow-up  spectroscopy observations  are consistent  with a  variable
Young Stellar  Object (YSO)  illuminating its surrounding  nebula.  In
this case the  source periodic variation along the  cycles produces an
unprecedented light echo in the different regions of the nebula.

\end{abstract}


\section{Introduction}
\label{sec:intro}

\begin{figure}
\centering
    \includegraphics[scale=0.65]{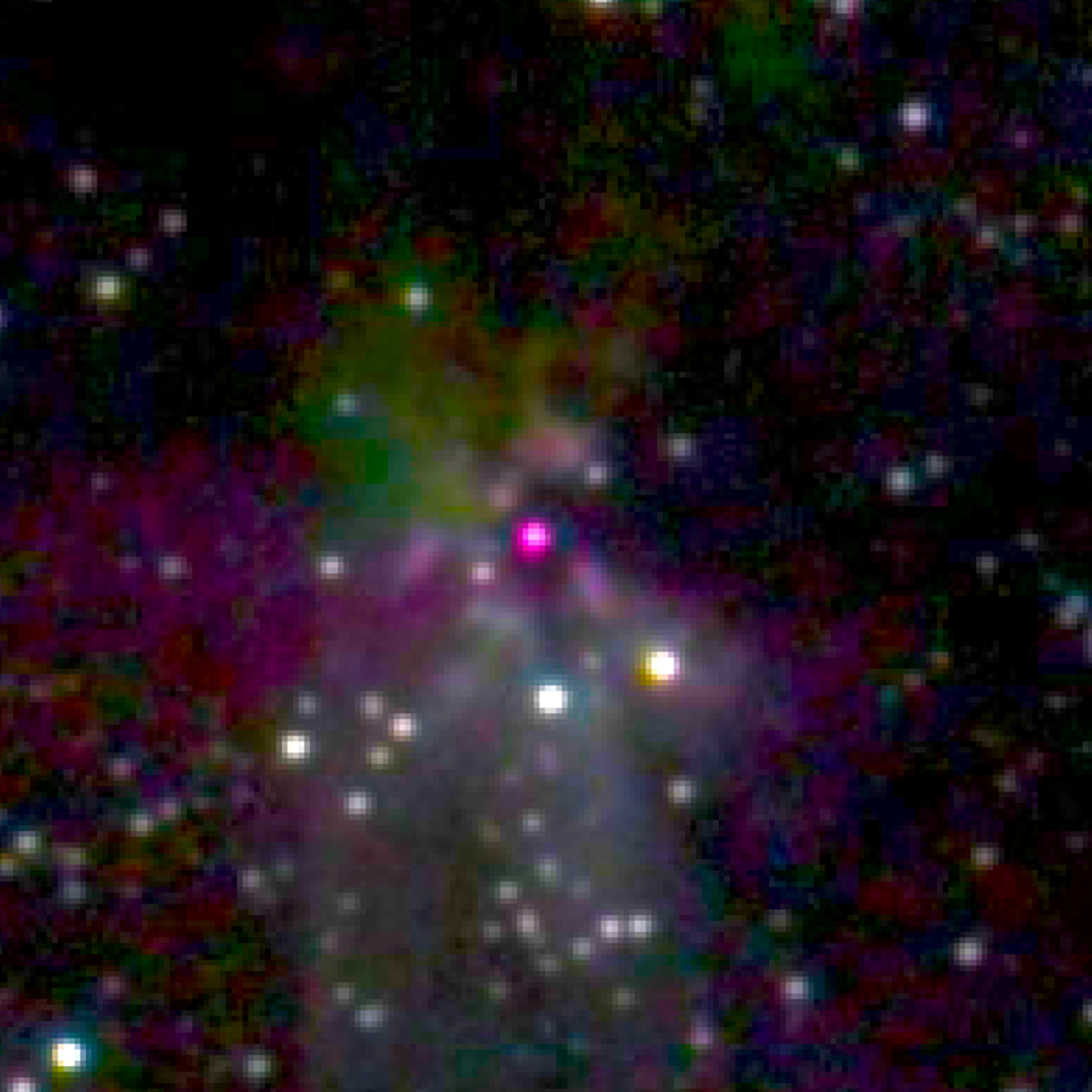}
    \caption{Composite  VVV-WIT-12 $K_{\rm  s}$-band image  made using
      three different epochs: years 2010 (red), 2011 (green), and 2012
      (blue).  This is the discovery  image for VVV-WIT-12, that shows
      the variation of the central source  as well as the nebula.  The
      image  is 67$''$  on a  side, centered  in VVV-WIT-12  at RA/DEC
      (J2000)  =  17:17:20.29,  $-$36:08:43.9   ($l,  b  =  350.7080$,
      $+1.0259$  deg),   and  oriented  along   Galactic  coordinates.
      Sources that are non-variable in  the near-IR should look white,
      while variable sources are colored differently. The point source
      was  brighter  in 2010  and  the  nebula changed  brightness  in
      different parts along the epochs.}
    \label{fig:image}
\end{figure}

\begin{figure*}[bt]
\centering
    \includegraphics[scale=0.63]{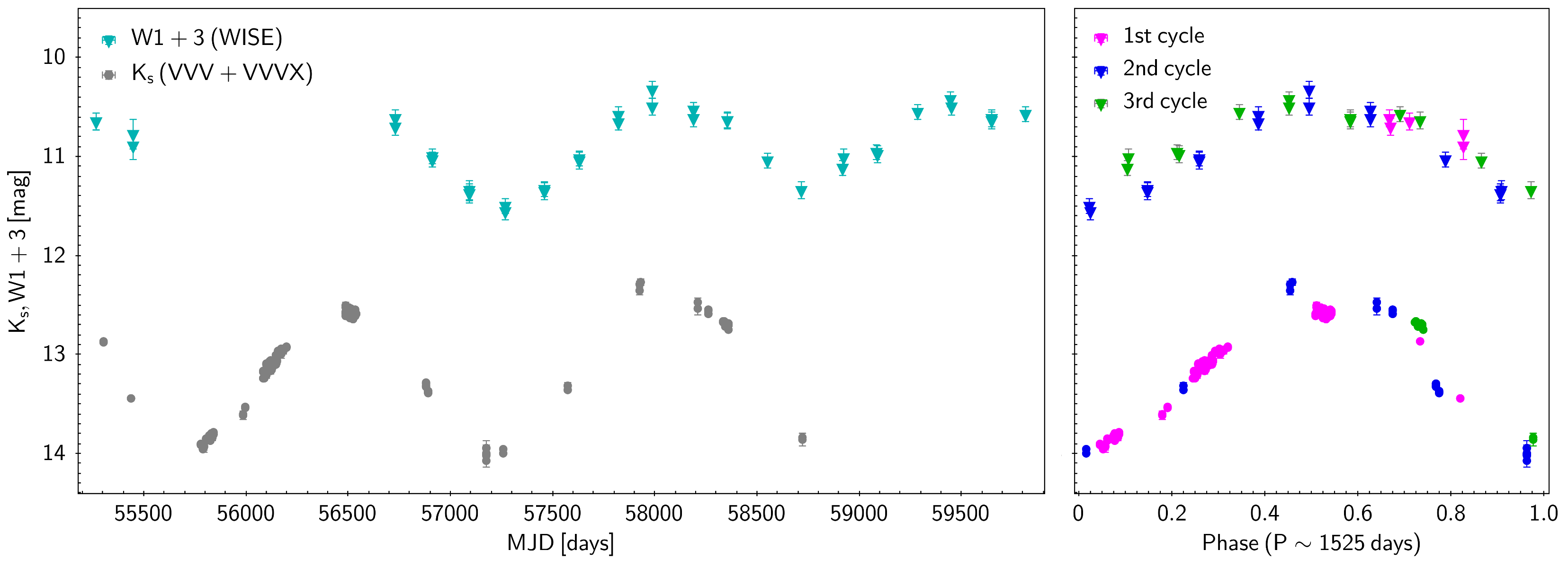}
    \caption{Left panel: Light curve of VVV-WIT-12 covering years 2010
      to 2022.  Grey points represent the K$_s$-band observations from
      VVV/VVVX,  and cyan  points represent  the W1-band  observations
      from  WISE  (arbitrarily  shifted   by  3  mag).   Right  panel:
      VVV-WIT-12  light   curves  phased  using   $P=1525$~days.   The
      different  cycles  are  color-coded  as  labelled  to  show  the
      amplitude changes. While the minimum  light seems to be the same
      in the two  cycles observed with the K$_s$-band,  the maximum is
      about  0.3  mag  brighter  in the  second  cycle,  indicating  a
      possible eruption.}
    \label{fig:lcurves}
\end{figure*}

A few  previously unclassified variable  sources that we  call ``WIT''
objects - short for What Is This?  - have been discovered by the VISTA
Variables     in      the     V\'ia     L\'actea      (VVV)     survey
\citep{2010NewA...15..433M}. These represent a wide variety of extreme
or rare  astrophysical phenomena, including  a very reddened  novae or
supernova         or          a         protostellar         collision
\citep[VVV-WIT-01;][]{2020MNRAS.492.4847L},  an   extragalactic  radio
source     violently     variable     in    the     near-IR     \citep
[VVV-WIT-04;][]{2019MNRAS.490.1171S},  another Tabby  star or  Mamajek
object  \citep[VVV-WIT-07;][]{2019MNRAS.482.5000S}  and a  giant  star
that blinked \citep[VVV-WIT-08][]{2021MNRAS.505.1992S}.

In year 2012 we started a pilot search for light echoes in the near-IR
images of  VVV survey. Light  echoes occur when the  interstellar dust
acts as a mirror reflecting the light of a very bright transient event
like    a    SN    \citep{2005MNRAS.357.1161P,    2006MNRAS.369.1949P,
  2008ApJ...681L..81R, 2011ApJ...732....2R}. We  made RGB color images
using observations in the $K_{\rm s}$-band acquired in three different
epochs from 2010 to 2012.  Very few candidates were detected, probably
because  the timespan  was short.   However, one  of these  candidates
showed a nebula with changing colors, around a strongly variable point
source that we named VVV-WIT-12 (see Fig.~\ref{fig:image}).

It  is not  clear if  VVV-WIT-12  produces a  light echo  or the  mere
illumination of the gaseous nebula by a strong variable source.  Since
this object  defies classification, we  labeled it  as one of  our WIT
objects.   This   discovery  is   relevant  because  the   Vera  Rubin
Observatory  (former  LSST) will  start  operations  soon, enabling  a
massive search  for light  echoes from  ancient SNe  in the  Milky Way
\citep{2022AJ....164..250L}.   Such  searches  have the  potential  to
discover  and monitor  the  evolution of  other  unusual objects  like
VVV-WIT-12.

\section{Observations and archive data}
\label{sec:obs}

VVV-WIT-12 is located  in the Galactic plane within VVV  tile b341, in
the direction  of the  bulge.  Its  Equatorial coordinates  are RA/DEC
(J2000)  =  17:17:20.29,   $-$36:08:43.9;  corresponding  to  Galactic
coordinates $l, b = 350.7080$,  $+1.0259$ deg.  The target is embedded
in a  compact nebula  that lies  within the  large RCW126  (GUM20) HII
region complex \citep{1976A&AS...25...25D}.

The parallax and VVV K$_s$-band light curve of VVV-WIT-12 was obtained
from VIRAC2  \citep{2018MNRAS.474.1826S,Smith_2023}. There is  a total
of 203 well  sampled epochs, spanning by 3418  days ($\sim$9.4 years),
from April 18 2010 to August 27 2019. The parallax is $\omega= 2.5 \pm
1.5$~mas, PM$_{\rm  RA}= -0.03 +/- 0.33$~mas  yr$^{-1}$, PM$_{\rm DEC}
=-0.30 \pm 0.33$~mas yr$^{-1}$.  Both  the parallax and proper motions
are consistent with zero, indicating a distant object. Based on VIRAC2
data,  \cite{2022MNRAS.509.2566M} classified  this  object  as a  long
period variable  (LPV), reporting  a period of  $P=292$~days.  Indeed,
our VIRAC2 light curve looks nearly sinusoidal, but with a much longer
period   of   $P=1525   \pm    30$~days,   resembling   a   LPV   (see
Fig.~\ref{fig:lcurves}).

Despite having acquired  many observations with the  other VVV filters
($ZYJH$), the  central point source  is not  detected in any  of these
shorter wavelengths in  order to establish if color  variations are in
synch with  the light curve.   However, these non-detection  allows to
estimate the  color limits using  the limiting magnitudes for  the VVV
tile  b341  \cite{2012A&A...537A.107S}.   In particular,  the  extreme
$(H-K_{\rm s}) > 5.0$~mag suggests  very high optical extinction, $A_V
\sim 100$~mag.  The mean extinction  towards this field, considering a
5$'$  FoV   is  $A_V=23.28$~mag  ($A_{K\rm  s}=2.69$~mag)   using  the
\cite{2011ApJ...737..103S}  extinction maps.   However, the  source is
indeed located  in a dense  and dark part of  the nebula, and  a total
extinction  $A_V   \sim  100$~mag  is  not   unrealistic.   Such  high
extinction is also confirmed by the spectral energy distribution (SED)
of the  point source,  shown in  Fig.~\ref{fig:spec}, which  is rather
extreme  (see  Appendix  \ref{sec:apeC}  for  details  about  the  SED
data). We  tried to  fit the  SED using a  variety of  stellar spectra
libraries   available   at   VOSA   \citep{2008A&A...492..277B},   and
\cite{2007ApJS..169..328R}   and  \cite{2015A&A...577A..42B}   stellar
models, however the results are  inconclusive since none of the models
resulted in a reasonably good fit to the data.  For instance, adopting
the  aforementioned extinction  and distance,  the SED  models suggest
this source is likely to be  a massive YSO.  Limiting ourselves to the
range $2~\mu m$ to $22~\mu m$ ($K_{\rm s}$ from VVV plus W1 to W4 from
WISE) the  SED can be  fitted to the  \cite{2007ApJS..169..328R} model
grid with a deeply  embedded, class 0 or class I  YSO model, but model
parameters should  be treated with  caution due  to the presence  of a
massive cloud core that dominates the far infrared SED and the unusual
nature of  the system.  The difficulty  in matching  the SED  with YSO
models arises because in the archival data the photometry was taken at
different  epochs,  thus  at  different  phases,  and  with  different
apertures (see Appendix \ref{sec:apeC}).

The VVV  images of  this region  show that  the vicinity  resembles an
active  star  formation region  (SFR),  being  rich with  interstellar
clouds, compact  cores, YSOs, and  young star clusters, as  example of
DB121    and    DB123    which     have    distances    measured    by
\cite{2008A&A...478..419S} as $D=1.6$~kpc.  Assuming the same distance
for VVV-WIT-12  and its nebula yields  a size of about  0.2~pc for the
whole  illuminated  nebular region  (30$''$).  While  this is  a  safe
assumption,   the  confirmation   of   the   distance  needs   further
observations.

\begin{figure}
\centering
    \includegraphics[scale=0.75]{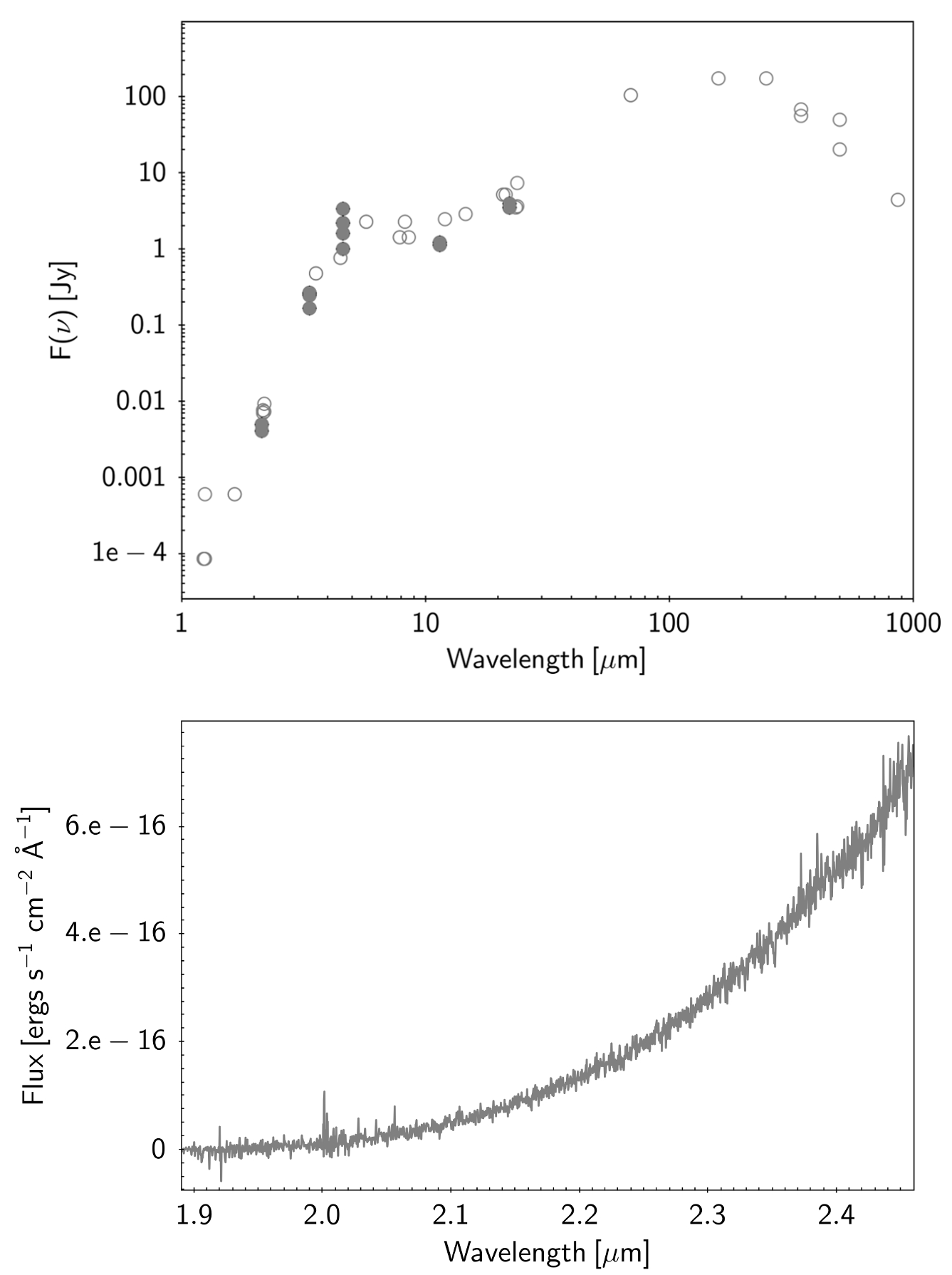}
    \caption{Top:  spectral energy  distribution  (SED) of  VVV-WIT-12
      covering  $1-1000~\mu m$,  obtained with  the Vizier  photometry
      viewer$^{\,\rm a}$. Solid circles mark  the data points from VVV
      and WISE within  the range $2~\mu m$ to $22~\mu  m$ (see Section
      \ref{sec:obs}).   Bottom:  near-IR  spectrum of  the  VVV-WIT-12
      stellar source, acquired with the TripleSpec spectrograph at the
      SOAR  telescope  in   Chile,  in  2023  June   11  (JD  2460106,
      corresponding  to $\phi=0.17$  cycles,  see the  right panel  of
      Fig.~\ref{fig:lcurves}).   This shows  that the  source is  very
      red,  with  a  featureless steadily  rising  continuum,  lacking
      strong       near-IR        emission       lines.\\       $^{\rm
        a}$\,\footnotesize{\url{http://vizier.cds.unistra.fr/vizier/sed/}}. \\ $\,\,$Data
      appearing in the SED are described in Appendix~\ref{sec:apeC}.}
    \label{fig:spec}
\end{figure}

\begin{figure}
\centering
    \includegraphics[scale=0.98]{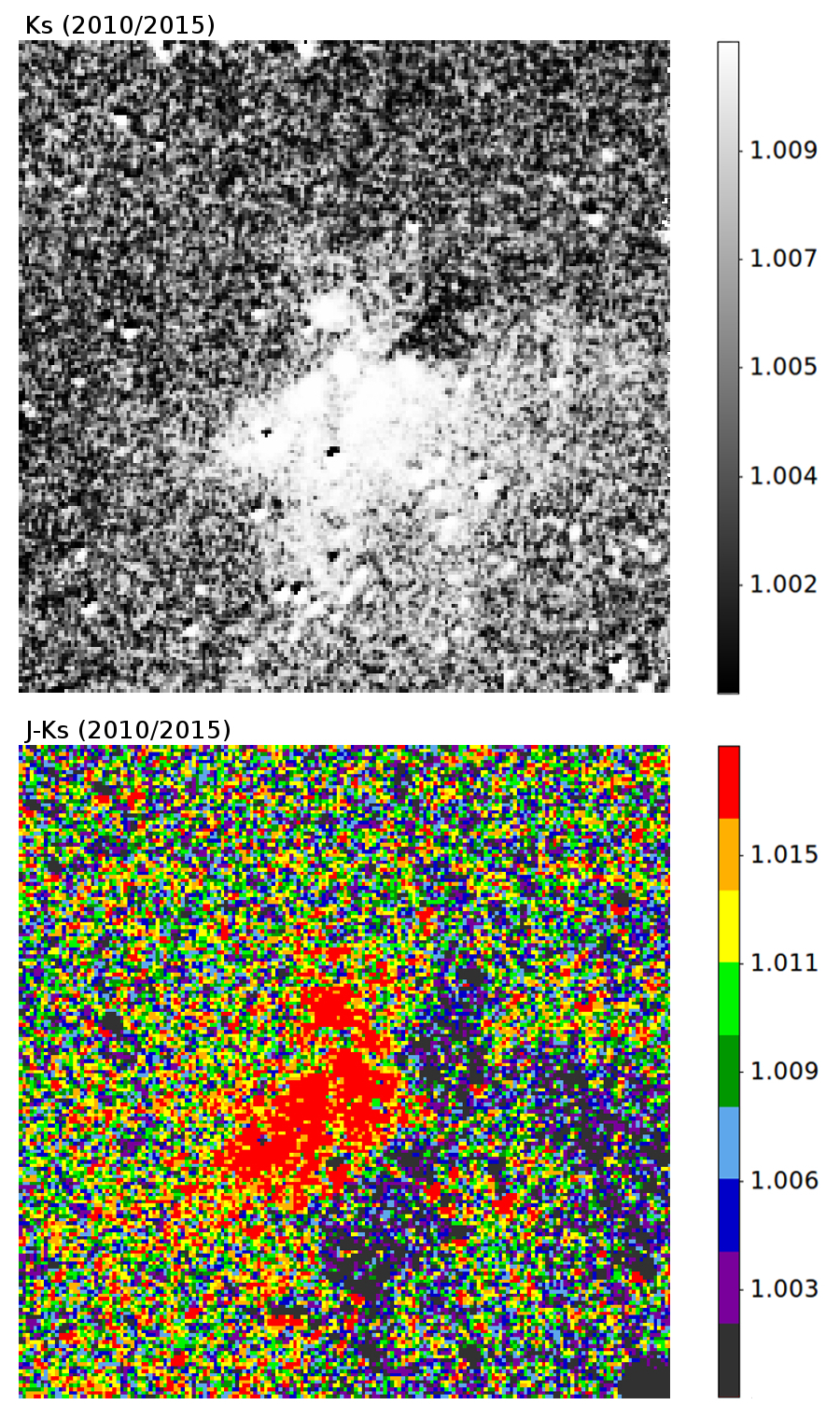}
    \caption{Top: Composite  $K_{\rm s}$-band image made  using a year
      2010 image  (with VVV-WIT-12 near phase  $\phi\sim 0.75$~cycles)
      divided  by a  year  2015 ($\phi\sim  0.05$~cycles). The  nebula
      varies  in  $K_{\rm s}$-band,  appearing  brighter  in the  first
      epoch. Bottom: Composite $J/K_{\rm s}$ image showing the spatial
      variation  of  the  near-IR  nebular  color.   Also,  the  color
      variations   cover   different   places  than   the   brightness
      variations. In  both panels  the field  of view  is 67$''$  on a
      side,  oriented  along   Galactic  coordinates,  with  longitude
      increasing to the left and  latitude towards the top, and covers
      the same field as Fig.~\ref{fig:image}. The color scale has been
      stretched to emphasize the changes.   A horizontal bar shows the
      relative intensity in each case.}
    \label{fig:nebula}
\end{figure}

Archive  search   at  the  VVV-WIT-12  position   revealed  additional
measurements  at longer  wavelengths.  In  particular, (NEO)WISE  data
\citep{2010AJ....140.1868W,2011ApJ...731...53M}   covers    a   longer
baseline than VVV, from March 12  2010 to August 21 2022, however with
a sparser cadence.  The W1-band  light curve presents a periodicity of
$\sim 4$~years with an amplitude $A_{W1} = 1.1$~mag, in agreement with
the VVV observations (see Fig.~\ref{fig:lcurves}). Such long period is
toward the upper  end of what is  seen in LPVs, and we  note that YSOs
can also  be periodic too.  The  amplitude is variable in  the $K_{\rm
  s}$-band  based on  3  cycles.   The first  and  second cycles  show
observed  peak-to-peak amplitudes  of  $A_{Ks}\sim1.4$ and  $1.7$~mag,
respectively.

\begin{figure*}
\centering
    \includegraphics[scale=0.7]{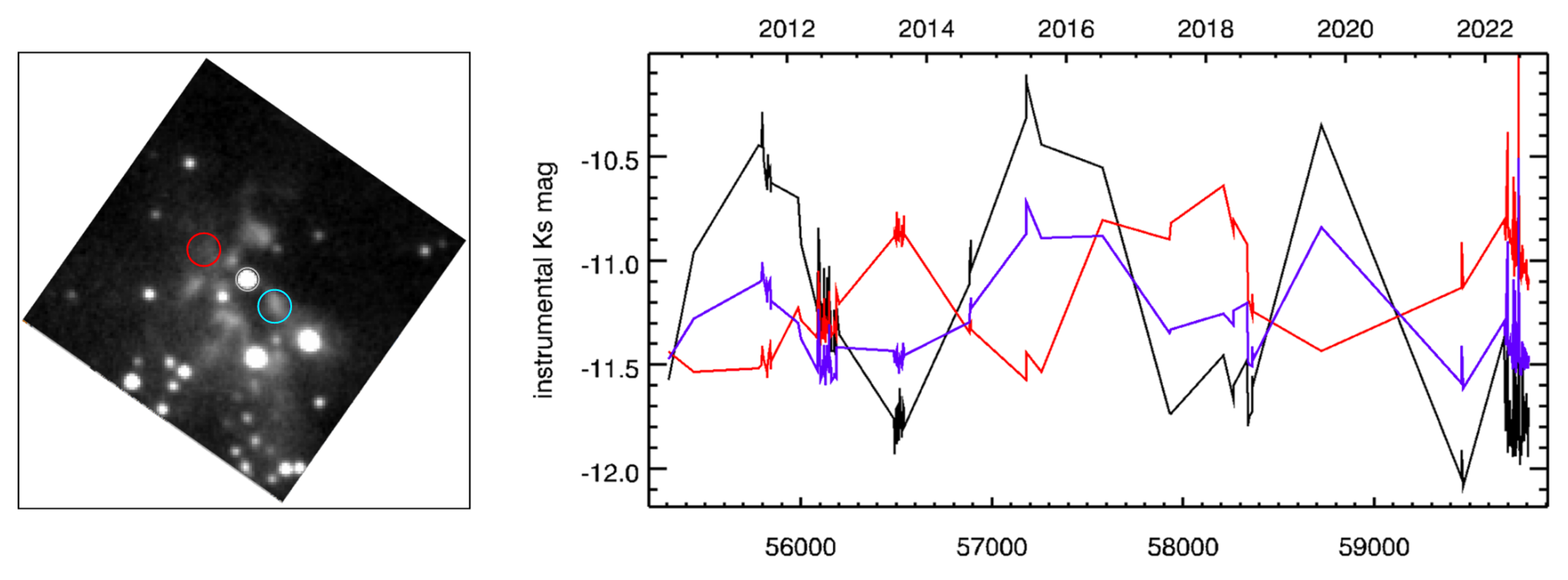}
    \caption{Variability of  the scattered  light from two  regions of
      the  VVV/VVVX  $K_{\rm  s}$  images.   Left:  the  emission  was
      integrated inside  the circular blue  and red masks.   A smaller
      mask was also placed on the object. The image orientation is the
      the  same  as  in  Figs.~\ref{fig:image}  and  \ref{fig:nebula}.
      Right: the resulting light curves for the three regions covering
      the same epochs as  Fig.~\ref{fig:lcurves}.  All curves show the
      same  periodicity  behavior,  but   the  red  curve  presents  a
      substantial lag with respect to the point source, while the blue
      curve is in phase with VVV-WIT-12.   Time in the x-axis is shown
      in years (top) and MJD (bottom).}
    \label{fig:echo}
\end{figure*}

The surrounding nebula changes its  brightness and $J-K_{\rm s}$ color
during the period of  observations (Fig.~\ref{fig:nebula}, see details
in Appendix \ref{sec:apeA}).   In year 2015 the  Northern part appears
redder than  in 2010,  while the opposite  occurred with  the Southern
part, which became bluer. Scattered  light from two regions located on
bright nebular patches at opposite  sides of VVV-WIT-12 shows the same
periodic behavior as the central source, confirming that the scattered
light is indeed  an echo of the source  (see Fig.~\ref{fig:echo}). The
red curve  shows a  substantial lag  (about half  a period)  which may
indicate that  it is caused by  back scattering off dust  more distant
than the source.  In  this case, if we take the  $2.1$~year lag as the
light travel  time from  the source  to the back  side of  the nebular
cavity, it  would yield a distance  of about $0.32$~pc.  On  the other
hand,  the blue  curve  is  in phase  with  central  source, which  is
interesting since  the forward scattered  light also has to  travel an
extra  distance   compared  to   the  direct  light.    However,  this
consideration only holds  for single scattering. Because  of the dense
environment  and the  large extinction  toward the  source, this  will
likely  not be  the case  here. Perhaps  the blue  region has  a lower
optical depth so that the light can more easily leak toward us.

The scale of the variations across the nebula is relevant, and we take
30$''$   as   a   representative   scale,   fully   contained   within
Fig.~\ref{fig:image}, that  covers 1.1$'$  on a  side.  We  use 30$''$
because this is the total size of the region that varies in brightness
and  color.  Assuming  a distance  $D=1.6$~kpc, 30$''$  corresponds to
$0.24$~pc.  In  this case,  the fluctuating region  would be  40 times
smaller  than  the  size  of   the  Orion  nebula  for  example.   The
light-travel time throughout a region of  this size would be about 0.8
yr. If the  source is more distant, e.g., located  within the bulge at
$D=8$~kpc, then this region would be larger, $30'' = 1.2$~pc.  This is
a large scale for brightness variations, considering that it cannot be
attributed  to gas  motions  because that  would require  superluminal
velocity.   Instead,  the preferred  explanation  is  that the  nebula
changes  as it  is being  illuminated by  the  variable central
source. We estimated  the range of luminosity variation  for the point
source using  the lowest and highest  values of fluxes in  the NIR/MIR
bands, assuming the distance of  1.6 kpc and no extinction correction.
This  yields  $\Delta   L  =  657  \pm  71~L_{\odot}$   and  $800  \pm
56~L_{\odot}$, where the errors were  derived by using the photometric
errors as lower/upper bounds.

Spectroscopic follow-up of the central source (VVV-WIT-12) was secured
on  June 11  2023  (JD 2460106,  corresponding  to $\phi=0.17$  cycles
according to the  ephemeris used to phase fold  our light-curves) with
the TripleSpec spectrograph\footnote{TripleSpec  instrument covers the
  wavelength range $\sim 0.80 - 2.47~\mu  m$ with a resolving power of
  $R \sim 3500$.  The spectra were reduced using a modified version of
  the IDL-based SpexTool and the  SOAR-TS4 pipeline was used to remove
  the   telluric  lines   by  conducting   an  A-B   position  pattern
  \citep{2004PASP..116..362C}.   The star  HIP 86098  was used  as the
  telluric standard for  the flux calibration.} at  the SOAR Telescope
\citep{2004SPIE.5492.1295W,2008SPIE.7014E..0XH}.      The     spectrum
(Fig.~\ref{fig:spec})  shows  that the  source  is  very red,  with  a
featureless steadily rising continuum, lacking strong near-IR emission
lines.  The  feature seen at  $\sim2.0~\mu m$  is the residual  of the
removal of a very strong telluric line.  Such featureless but very red
spectra   are  typical   of   some  YSOs   \citep{2020MNRAS.495.3614C,
  2022MNRAS.513.1015G}.   However,   a  very  dusty  Mira   cannot  be
discarded, as they may also exhibit a similar spectral behaviour.

\section{Discussion}
\label{sec:disc}

As with most VVV-WIT objects discovered,  we still do not know what is
this.   Not only  the source  is  variable, but  also the  surrounding
nebulosity changes  brightness and shape.   Given the long  period and
the   classification  as   an   LPV,  initially   we  considered   two
possibilities:  (i)  an  LPV   that  illuminates  the  encircling  ISM
inhomogeneously through thick  dust clouds in its  atmosphere, or (ii)
an LPV that is coming out from behind an unrelated interstellar cloud.

The  LPV hypothesis  is discarded  by examining  the magnitudes.   The
expected  magnitude for  an  LPV with  $P=1525$~  days is  $M_{Ks}\sim
-11$~mag  \citep[e.g.,][]{2017EPJWC.15207001M}.   If   the  object  is
located  at  $D=1.6$~kpc,  its   distance  modulus  would  be  $m-M_0=
11.02$~mag.   Therefore its  apparent unreddened  magnitude should  be
about $0$~mag,  which is inconsistent  with the observed  magnitude of
$K_{\rm s}$=13.2~mag. A  possible exception might have a  rare case of
J~type Carbon  star, that are  fainter than other AGB  stars, reaching
$M{_K}  \sim  -2.5$~mag \citep{2020A&A...633A.135A}.   However,  these
objects are rather blue, inconsistent with the observed colors and the
spectral energy distribution that is rising to the near-IR.

We also  note that the  observed geometry  for the nebular  changes is
inconsistent with the spherically symmetric  picture that a light echo
would  produce, as  seen for  example in  historical SNe  remnants and
$\eta$\,\,Carina \citep{2005Natur.438.1132R}.  This  suggests that the
dust configuration surrounding VVV-WIT-12 causes a light echo which is
by  far  more complex  than  those  of  SN.   In principle,  the  dust
distribution  could be  characterized based  on the  phase lag  of the
scattered light, but this detailed 3D  modeling is beyond the scope of
this paper.

Considering now that VVV-WIT-12 is a YSO because of its luminosity and
spectrum as  well as  the long  term oscillatory  pattern, as  seen in
other   YSOs   in   the  mid-infrared   \citep[e.g.,][and   references
  therein]{2021ApJ...920..132P},   attributed   to   either   variable
extinction or  a cyclical change  in accretion  rate, there are  a few
other alternatives  to explain  the observations:  (i) the  light echo
from a periodic  YSO, or (ii) the progressive illumination  of the ISM
by  the precessing  disk  of a  YSO,  or (iii)  an  orbiting clump  of
material  that blocks  the light  path of  the variable  source.  Even
though  at present  the  YSO  option remains  as  the most  reasonable
hypothesis,   we  are   unable   to  confirm   this  without   further
observations.

\section{Conclusions}
\label{sec:conclusions}

We  have  discovered  VVV-WIT-12,  a  very  reddened  ($A_V>100$  mag)
periodic variable  source with  $P=1525$ days, that  has a  smooth red
continuum  and  that  is  located  inside a  nebula  that  changes  in
brightness and color with time. The real nature of the source is still
unknown, but based on the  extensive near-IR imaging and spectroscopic
observations, our preferred explanation is a periodically variable YSO
with a light echo.

In particular, massive  light echo searches are being  planned for the
Vera   C.   Rubin   Observatory  \citep{2022AJ....164..250L},   taking
advantage of  a median of 815  images scheduled for every  position of
the  sky, reaching  $g  \sim  24.5$ mag,  $r  \sim  24.0$ mag.   These
searches will  use artificial  intelligence tools like  automated deep
convolutional  neural  networks  to  make many  important  light  echo
discoveries  from past  SNe,  especially across  the  Milky Way  plane
\citep{2005Natur.438.1132R}.   Our work  suggests  that these  massive
light echo explorations  may be contaminated by  variable objects like
the one presented here.  However, these light echo searches will allow
to discover and  classify VVV-WIT-12 analogs, and to  study their time
evolution.  Also,  not less important,  these searches would  open the
door for other serendipitous discoveries.

In conclusion,  VVV-WIT-12 appears to be  a very rare long  period YSO
that acts as a cosmic beacon flashing its surrounding nebula.  This is
worthy of further study, and suggested future observations are near-IR
spectroscopy of the nebula to  confirm its nature, and more continuous
monitoring of the variability of this interesting object.

\begin{acknowledgments}
We gratefully acknowledge  the use of data from the  ESO Public Survey
program IDs 179.B-2002  and 198.B-2004 taken with  the VISTA telescope
and data  products from the  Cambridge Astronomical Survey  Unit. This
publication makes  use of data  products from the  Wide-field Infrared
Survey  Explorer,  which is  a  joint  project  of the  University  of
California, Los Angeles, and  the Jet Propulsion Laboratory/California
Institute of Technology, funded by  the National Aeronautics and Space
Administration. D.M.  gratefully acknowledges  support from the Center
for Astrophysics  and Associated Technologies  CATA by the  ANID BASAL
projects ACE210002 and FB210003, by Fondecyt Project No.  1220724, and
by  CNPq/Brazil through  project 350104/2022-0.   R.K.S.  acknowledges
support   from   CNPq/Brazil   through  projects   308298/2022-5   and
350104/2022-0.  ZG  is supported  by  the  ANID Fondecyt  Postdoctoral
program No.  3220029.  ZG acknowledges  support by ANID, -- Millennium
Science Initiative Program -- NCN19\_171.  The work of FN is supported
by NOIRLab,  which is managed  by the Association of  Universities for
Research in  Astronomy (AURA) under  a cooperative agreement  with the
National  Science Foundation.  CM acknowledges  support from  the UK's
Science and Technology Facilities Council (ST/S505419/1).

\end{acknowledgments}

\appendix

\section{Analysis of the nebula}
\label{sec:apeA}
 
To  inspect  the  variability  of   the  scattered  light  across  the
surrounding nebula, we analyze the  VVV/VVVX images taken at different
epochs.  In  a first step,  we selected $K_{\rm s}$-band  images taken
near to  maximum/minimum of  the VVV-WIT-12 light-curve,  according to
the $P=1525$~ephemeris:  a $K_{\rm s}$-band  image taken on  April 18,
2010, at  phase $\phi\sim 0.75$;  and a $K_{\rm  s}$-band image
from  August  15, 2015,  at  phase  $\phi\sim 0.05$.   Then  we
divided  the 2010  image by  the  2015, creating  a composite  $K_{\rm
  s}$-band   image   with   the    differences.    As   presented   in
Fig.~\ref{fig:nebula},  the nebula  changes  in brightness,  appearing
brighter in  the first  epoch, when  the VVV-WIT-12 is closer  to the
maximum ($\phi\sim 0.75$).  We then  repeated the same steps as
above using $J$ and $K_{\rm s}$-band images taken in sequence at those
nights  (in 2010  and  2015).   A composite  $J/K_{\rm  s}$ image  was
created by dividing the $J$-band  by the $K_{\rm s}$-band, and finally
dividing the 2010 $J/K_{\rm s}$ image  by the 2015 $J/K_{\rm s}$. This
shows the spatial variation of  the near-IR nebular color, which cover
different  places  than  the  brightness variations.   After  all  the
divisions,  the  net differences  both  in  brightness and  color  are
$\sim1-2\%$,   but    consistent   across   all   the    nebula.    In
Fig.~\ref{fig:nebula} the color scale  has been stretched to emphasize
the changes.

Since  our  first  analysis  using  two  epoch  showed  variations  in
brightness  and color  across the  nebula,  we made  use of  a set  of
$K_{\rm  s}$-band images  taken along  the VVV/VVVX  campaign, ranging
from from April  2010 to August 2019. We selected  two regions located
on bright  nebular patches at  opposite sides  of the point  source to
place two circular  masks of 6 pixel radius in  order to integrate the
emission, after subtracting  the background. A third  and smaller mask
of 4 pixel radius was also placed  on the position of the point source
(VVV-WIT-12),  for  simultaneous monitoring  (see  the  left panel  of
Fig.~\ref{fig:echo}).  The  extracted magnitudes  (instrumental, i.e.,
not photometrically calibrated) were combined with the MJD to create a
resulting  light   curve  for   both  masks   placed  on   the  nebula
simultaneously  with  the  central  point source.  In  extracting  the
magnitudes,  the  uncertainty is  in  the  range of  $\sigma_{\rm  Ks}
\approx 0.2$~mag.  The light curves for the two regions (blue and red)
were arbitrarily shifted to approximately  match the mean level of the
point  source (black  light curve),  as presented  in the  right panel
Fig.~\ref{fig:echo}.  All  curves show the same  periodicity behavior,
but the red curve presents a substantial lag with respect to the point
source, while the blue curve is in phase with VVV-WIT-12.

\section{VIRAC2  $K_{\rm s}$-band light-curve}
\label{sec:apeB}

Here we present all the VIRAC2 photometric measurements of the central
source  VVV-WIT-12  used  to  build   the  light  curve  presented  in
Fig.~\ref{fig:lcurves}.   These data  listed in  Table 1  comprise 203
$K_{\rm s}$-band  data-points.  The  source is relatively  bright, and
the typical photometric errors are very small ($\sim 0.02$ mag).

\startlongtable
\begin{deluxetable}{cc}
\tablecaption{VVV-WIT-12 $K_{\rm  s}$-band variabiliity data}
\tablehead{
\colhead{MJD} & \colhead{$K_{\rm s}$-band} \\ 
\colhead{(days)} & \colhead{(mag)}
} 
\startdata
55305.31399078   &  12.882$\pm$0.014 \\   
55305.31435675   &  12.872$\pm$0.012 \\   
55437.11724679   &  13.440$\pm$0.015 \\   
55437.11768706   &  13.448$\pm$0.010 \\   
55779.09640132   &  13.906$\pm$0.014 \\   
55779.09682943   &  13.911$\pm$0.019 \\   
55791.14535418   &  13.954$\pm$0.013 \\   
55791.14572990   &  13.956$\pm$0.017 \\   
55796.03104124   &  13.936$\pm$0.053 \\   
55796.03145355   &  13.923$\pm$0.046 \\   
55804.09243890   &  13.855$\pm$0.013 \\   
55804.09283343   &  13.871$\pm$0.016 \\   
55822.03137893   &  13.835$\pm$0.014 \\   
55822.03181434   &  13.839$\pm$0.015 \\   
55822.04318502   &  13.851$\pm$0.015 \\   
55822.04358429   &  13.849$\pm$0.018 \\   
55823.01212277   &  13.840$\pm$0.021 \\   
55823.01254251   &  13.814$\pm$0.017 \\   
55826.99329101   &  13.869$\pm$0.021 \\   
55826.99370021   &  13.869$\pm$0.026 \\   
55829.07088450   &  13.821$\pm$0.028 \\   
55829.07126727   &  13.802$\pm$0.015 \\   
55838.01804222   &  13.821$\pm$0.018 \\   
55838.01843721   &  13.837$\pm$0.028 \\   
55842.00936837   &  13.786$\pm$0.016 \\   
55842.00976777   &  13.809$\pm$0.015 \\   
55842.02231214   &  13.797$\pm$0.012 \\   
55842.02272759   &  13.788$\pm$0.015 \\   
55843.00842881   &  13.807$\pm$0.022 \\   
55843.00881339   &  13.812$\pm$0.017 \\   
55984.32586264   &  13.614$\pm$0.041 \\   
55984.32623896   &  13.605$\pm$0.031 \\   
55999.28517953   &  13.537$\pm$0.019 \\   
55999.28555562   &  13.526$\pm$0.019 \\   
56084.20270002   &  13.240$\pm$0.010 \\   
56084.20316717   &  13.238$\pm$0.013 \\   
56086.22350249   &  13.168$\pm$0.011 \\   
56086.22399201   &  13.179$\pm$0.009 \\   
56091.17856161   &  13.230$\pm$0.021 \\   
56091.17908462   &  13.238$\pm$0.023 \\   
56099.27448288   &  13.189$\pm$0.027 \\   
56099.27495279   &  13.207$\pm$0.027 \\   
56101.13757731   &  13.148$\pm$0.013 \\   
56101.13805174   &  13.157$\pm$0.022 \\   
56102.21649852   &  13.089$\pm$0.014 \\   
56102.21698412   &  13.100$\pm$0.018 \\   
56112.07114756   &  13.159$\pm$0.017 \\   
56112.07158345   &  13.157$\pm$0.017 \\   
56113.98610091   &  13.136$\pm$0.027 \\   
56113.98648592   &  13.133$\pm$0.015 \\   
56114.98102135   &  13.072$\pm$0.014 \\   
56114.98143780   &  13.074$\pm$0.015 \\   
56115.98474485   &  13.110$\pm$0.023 \\   
56115.98515314   &  13.105$\pm$0.024 \\   
56120.04092490   &  13.069$\pm$0.018 \\   
56120.04145048   &  13.067$\pm$0.012 \\   
56121.12187599   &  13.159$\pm$0.029 \\   
56121.12243938   &  13.167$\pm$0.029 \\                                 
56121.97479523   &  13.061$\pm$0.019 \\                               
56121.97517811   &  13.074$\pm$0.020 \\     
56123.01474560   &  13.164$\pm$0.026 \\     
56123.01514024   &  13.145$\pm$0.022 \\     
56124.06170449   &  13.121$\pm$0.014 \\     
56124.06210611   &  13.154$\pm$0.013 \\     
56125.04530593   &  13.123$\pm$0.023 \\     
56125.04569999   &  13.110$\pm$0.019 \\     
56126.15988966   &  13.144$\pm$0.018 \\     
56126.16028760   &  13.145$\pm$0.013 \\     
56129.13860467   &  13.103$\pm$0.014 \\     
56129.13901743   &  13.100$\pm$0.013 \\     
56132.03022292   &  13.097$\pm$0.015 \\     
56132.03062791   &  13.107$\pm$0.014 \\     
56133.02148136   &  13.086$\pm$0.011 \\     
56133.02186186   &  13.087$\pm$0.015 \\     
56134.01105021   &  13.099$\pm$0.043 \\     
56134.01146314   &  13.065$\pm$0.038 \\     
56144.99923930   &  13.076$\pm$0.012 \\     
56144.99963643   &  13.069$\pm$0.013 \\     
56145.06523432   &  13.081$\pm$0.016 \\     
56145.06561676   &  13.078$\pm$0.013 \\     
56145.14724240   &  13.063$\pm$0.014 \\     
56145.14764321   &  13.068$\pm$0.014 \\     
56146.15497943   &  13.098$\pm$0.024 \\     
56146.15537408   &  13.092$\pm$0.019 \\     
56147.08220646   &  13.013$\pm$0.017 \\     
56147.08260093   &  13.004$\pm$0.016 \\     
56151.16657051   &  13.063$\pm$0.038 \\     
56151.16695293   &  13.066$\pm$0.026 \\     
56158.03083614   &  12.962$\pm$0.012 \\     
56158.03126193   &  12.981$\pm$0.018 \\     
56158.06806029   &  12.987$\pm$0.012 \\     
56158.06845358   &  12.976$\pm$0.018 \\     
56171.97870177   &  12.989$\pm$0.015 \\     
56171.97914848   &  12.998$\pm$0.016 \\     
56172.05053745   &  12.943$\pm$0.012 \\     
56172.05096107   &  12.953$\pm$0.012 \\     
56173.00648989   &  12.995$\pm$0.037 \\     
56173.00688473   &  12.984$\pm$0.029 \\     
56173.10230482   &  12.966$\pm$0.018 \\     
56173.10269927   &  12.957$\pm$0.016 \\     
56184.02361554   &  12.959$\pm$0.019 \\     
56184.02402129   &  12.948$\pm$0.014 \\     
56185.08016956   &  12.966$\pm$0.026 \\     
56185.08055668   &  12.963$\pm$0.033 \\     
56197.99238377   &  12.934$\pm$0.017 \\     
56197.99277893   &  12.925$\pm$0.011 \\     
56486.05191910   &  12.613$\pm$0.010 \\     
56486.05232836   &  12.588$\pm$0.019 \\     
56488.21161727   &  12.582$\pm$0.022 \\     
56488.21203659   &  12.567$\pm$0.021 \\     
56489.09005584   &  12.526$\pm$0.026 \\     
56489.09049970   &  12.501$\pm$0.026 \\     
56497.12600446   &  12.566$\pm$0.018 \\     
56497.12642201   &  12.547$\pm$0.017 \\     
56501.04456267   &  12.529$\pm$0.015 \\
56501.04497836   &  12.564$\pm$0.017 \\ 
56502.07928420   &  12.528$\pm$0.008 \\ 
56502.07970634   &  12.522$\pm$0.017 \\ 
56511.00270899   &  12.563$\pm$0.014 \\ 
56511.00311435   &  12.638$\pm$0.013 \\ 
56512.05194052   &  12.545$\pm$0.022 \\ 
56512.05239296   &  12.549$\pm$0.012 \\ 
56512.15321218   &  12.533$\pm$0.011 \\ 
56512.15364685   &  12.537$\pm$0.010 \\ 
56513.00073429   &  12.585$\pm$0.013 \\ 
56513.00114849   &  12.578$\pm$0.022 \\ 
56513.09750733   &  12.582$\pm$0.011 \\ 
56513.09793096   &  12.597$\pm$0.012 \\ 
56514.02949870   &  12.559$\pm$0.025 \\ 
56514.02992590   &  12.584$\pm$0.039 \\ 
56522.08995637   &  12.643$\pm$0.015 \\ 
56522.09035346   &  12.638$\pm$0.014 \\ 
56522.98801717   &  12.580$\pm$0.016 \\ 
56522.98842534   &  12.594$\pm$0.013 \\ 
56523.08986704   &  12.610$\pm$0.014 \\ 
56523.09029174   &  12.598$\pm$0.016 \\ 
56524.10995656   &  12.594$\pm$0.017 \\ 
56524.11036344   &  12.595$\pm$0.015 \\ 
56525.00600829   &  12.591$\pm$0.009 \\ 
56525.00643632   &  12.592$\pm$0.015 \\ 
56525.10749755   &  12.597$\pm$0.012 \\ 
56525.10789068   &  12.572$\pm$0.016 \\ 
56526.01769841   &  12.608$\pm$0.013 \\ 
56526.01810583   &  12.621$\pm$0.014 \\ 
56526.13782058   &  12.621$\pm$0.008 \\ 
56526.13820470   &  12.614$\pm$0.010 \\ 
56527.03777189   &  12.621$\pm$0.021 \\ 
56527.03819244   &  12.620$\pm$0.016 \\ 
56535.99104194   &  12.546$\pm$0.021 \\ 
56535.99147113   &  12.567$\pm$0.015 \\ 
56536.06409385   &  12.609$\pm$0.023 \\ 
56536.06451513   &  12.608$\pm$0.015 \\ 
56536.12114000   &  12.592$\pm$0.014 \\ 
56536.12153927   &  12.571$\pm$0.018 \\ 
56536.98901042   &  12.574$\pm$0.019 \\ 
56536.98944144   &  12.573$\pm$0.025 \\ 
56537.03435051   &  12.561$\pm$0.010 \\ 
56537.03477811   &  12.566$\pm$0.014 \\ 
56537.08520569   &  12.567$\pm$0.016 \\ 
56537.08559973   &  12.572$\pm$0.014 \\ 
56538.05154343   &  12.588$\pm$0.010 \\ 
56538.05195734   &  12.592$\pm$0.013 \\ 
56881.05502514   &  13.285$\pm$0.014 \\ 
56881.05544981   &  13.291$\pm$0.018 \\ 
56881.15602101   &  13.315$\pm$0.017 \\ 
56881.15640352   &  13.327$\pm$0.010 \\ 
56890.10222876   &  13.373$\pm$0.020 \\ 
56890.10264517   &  13.380$\pm$0.016 \\ 
56890.16151307   &  13.389$\pm$0.014 \\ 
56890.16192479   &  13.382$\pm$0.018 \\
57178.28608468   &  13.995$\pm$0.037 \\
57178.28647014   &  13.951$\pm$0.035 \\  
57178.33491099   &  14.012$\pm$0.050 \\  
57178.33531670   &  14.070$\pm$0.059 \\  
57178.37275000   &  14.019$\pm$0.058 \\  
57178.37315627   &  13.947$\pm$0.069 \\  
57259.00702552   &  13.954$\pm$0.011 \\  
57259.00838138   &  13.994$\pm$0.014 \\  
57576.21099736   &  13.311$\pm$0.033 \\  
57576.21138467   &  13.357$\pm$0.022 \\  
57926.32108585   &  12.290$\pm$0.034 \\  
57926.32144990   &  12.357$\pm$0.046 \\  
57933.31940204   &  12.267$\pm$0.035 \\  
57933.31975150   &  12.270$\pm$0.033 \\
58210.33530629   &  12.535$\pm$0.061 \\
58210.33569310   &  12.473$\pm$0.045 \\
58263.23436190   &  12.585$\pm$0.011 \\
58263.23471955   &  12.588$\pm$0.018 \\
58264.18160133   &  12.550$\pm$0.017 \\
58264.18197642   &  12.553$\pm$0.022 \\
58335.02426649   &  12.664$\pm$0.015 \\
58335.02466130   &  12.673$\pm$0.013 \\
58339.08740732   &  12.678$\pm$0.015 \\
58339.08777100   &  12.667$\pm$0.017 \\
58346.06097997   &  12.717$\pm$0.022 \\
58346.06134360   &  12.713$\pm$0.016 \\
58359.00003648   &  12.701$\pm$0.019 \\
58359.00039320   &  12.682$\pm$0.015 \\
58360.02341993   &  12.700$\pm$0.020 \\
58360.02382860   &  12.709$\pm$0.026 \\
58362.99737906   &  12.752$\pm$0.017 \\
58723.07908335   &  13.842$\pm$0.045 \\
58723.07946611   &  13.857$\pm$0.060 
\enddata
\end{deluxetable}

\newpage

\section{Archival data from Vizier SED}
\label{sec:apeC}

Here we  present archival data  from VizieR Photometry  Viewer (Vizier
SED)\footnote{http://vizier.cds.unistra.fr/vizier/sed/},  presented in
Fig.~\ref{fig:spec}.   Wavelength coverage  ranges from  $1~\mu m$  to
$1000~\mu m$.  In  the near-IR we allowed  1~arcsec maximum separation
from  VVV-WIT-12   coordinates  as   RA/DEC  (J2000)   =  17:17:20.29,
$-$36:08:43.9.   In  the mid-IR  and  far-IR  we relaxed  the  allowed
separation  to  5~arcsec  because  of  the  lower  resolution  of  the
instruments at these  longer wavelengths. The archival  data came from
several different catalogues under the  CDS Vizier database.  The full
information and metadata for each  catalogue can be found by accessing
the  VizieR  Photometry  Viewer   service${^2}$  with  the  VVV-WIT-12
coordinates and then selecting the  desired catalogue. We caution that
the  archival data  have  been taken  at different  epochs  - thus  at
different  phases  for  VVV-WIT-12  -  as well  as  depending  on  the
aperture/spatial  resolution the  flux measurements  can include  some
contamination from the surrounding nebula to the central source.

\startlongtable
\begin{deluxetable*}{llrcccl}
\tablecaption{Data from VizieR Photometry  Viewer presented in the SED
  of Fig.~\ref{fig:spec}.}
\tablehead{
\colhead{Filter}  & \colhead{Telescope}     & \colhead{Frequency} & \colhead{Flux}  & \colhead{Flux Error}  & \colhead{Separation} & \colhead{Catalogue}  \\ 
\colhead{(name) } & \colhead{            }  & \colhead{(GHz)}     & \colhead{(Jy)}  & \colhead{(Jy)}        & \colhead{(arcsec)}     & \colhead{(Vizier code)}
} 
\startdata
J       & 2MASS          &  241960   &  8.33E-5 &  ""       &  0.80  &   I/297/out                \\
J       & 2MASS          &  239830   &  8.50E-5 &  ""       &  0.03  &   II/246/out               \\
J       & 2MASS          &  239830   &  5.88E-4 &  ""       &  0.03  &   II/246/out               \\
H       & 2MASS          &  181750   &  5.94E-4 &  ""       &  0.03  &   I/297/out                \\
Ks      & VISTA          &  140500   &  0.00485 &  2.0E-5   &  0.24  &   II/376/vvv4              \\
Ks      & VISTA          &  140500   &  0.00491 &  2.0E-5   &  0.27  &   II/348/vvv2              \\
Ks      & VISTA          &  140500   &  0.00491 &  2.0E-5   &  0.18  &   II/337/vvv1              \\
Ks      & VISTA          &  140500   &  0.00398 &  0.00107  &  0.27  &   II/364/virac             \\
Ks      & 2MASS          &  138550   &  0.00748 &  5.1E-4   &  0.55  &   J/AJ/149/64/archive      \\
Ks	& 2MASS          &  138550   &  0.00687 &  4.7E-4   &  0.55  &   J/AJ/149/64/archive      \\
Ks      & 2MASS          &  136890   &  0.00904 &  1.9E-4   &  0.03  &   II/246/out               \\
Ks      & 2MASS          &  136890   &  0.00724 &  4.9E-4   &  0.03  &   II/246/out               \\
W1      & WISE           &   89490   &   0.166  &   0.002   &  1.14  &   II/365/catwise           \\
W1      & WISE           &   89490   &   0.241  &   0.0     &  1.02  &   II/363/unwise            \\
W1      & WISE           &   89490   &   0.255  &   0.005   &  0.80  &   II/328/allwise           \\
W1      & WISE           &   89490   &   0.261  &   0.006   &  0.54  &   II/311/wise              \\
3.6     & Spitzer/IRAC   &   84449   &   0.472  &   0.025   &  0.19  &   II/293/glimpse           \\
4.5     & Spitzer/IRAC   &   66724   &   0.763  &   0.095   &  0.19  &   II/293/glimpse           \\
W2      & WISE           &   65172   &   1.0    &   0.02    &  1.14  &   II/365/catwise           \\
W2      & WISE           &   65172   &   1.58   &   0.0     &  1.02  &   II/363/unwise            \\
W2      & WISE           &   65172   &   3.32   &   0.23    &  0.80  &   II/328/allwise           \\
W2      & WISE           &   65172   &   2.21   &   0.07    &  0.54  &   II/311/wise              \\
5.8     & Spitzer/IRAC   &   52311   &   2.32   &   0.06    &  0.19  &   II/293/glimpse           \\
8.0     & Spitzer/IRAC   &   38083   &   1.45   &   0.08    &  0.19  &   II/293/glimpse           \\
A       & MSX            &   36207   &   2.27   &   0.09    &  0.48  &   V/114/msx6\_gp           \\
S9W     & AKARI          &   34819   &   1.4    &   0.06    &  1.65  &   II/297/irc               \\
W3      & WISE           &   25934   &   1.23   &   0.02    &  0.80  &   II/328/allwise           \\
W3      & WISE           &   25934   &   1.11   &   0.02    &  0.54  &   II/311/wise              \\
C       & MSX            &   24715   &   2.5    &   0.14    &  0.48  &   V/114/msx6\_gp           \\
D       & MSX            &   20464   &   2.91   &   0.18    &  0.48  &   V/114/msx6\_gp           \\
$21~\mu m$    & Hi-GAL         &   14276   &   5.28   &   0.33    &  3.38  &   J/MNRAS/471/100/hcatalog \\
E       & MSX            &   14048   &   5.28   &   0.33    &  0.48  &   V/114/msx6\_gp            \\
W4      & WISE           &   13571   &   3.90   &   0.09    &  0.80  &   II/328/allwise           \\
W4      & WISE           &   13571   &   3.52   &   0.09    &  0.54  &   II/311/wise              \\
24      & Spitzer/MIPS   &   12663   &   3.53   &   0.0     &  0.77  &   II/368/sstsl2            \\
$ 24~\mu m$   & Hi-GAL         &   12491   &   7.43   &   0.1     &  3.38  &   J/MNRAS/471/100/hcatalog \\
$ 24~\mu m$   & Hi-GAL         &   12491   &   3.71   &   0.07    &  3.38  &   J/MNRAS/471/100/hcatalog \\
$ 70~\mu m$   & Hi-GAL         &  4282.8   &  105.0   &   5.0     &  3.38  &   J/MNRAS/471/100/hcatalog \\
$ 70~\mu m$   & Hi-GAL         &  4282.8   &  105.0   &   5.0     &  3.38  &   J/MNRAS/471/100/hcatalog \\
$160~\mu m$   & Hi-GAL         &  1873.7   &  173.0   &   3.0     &  3.38  &   J/MNRAS/471/100/hcatalog \\
$250~\mu m$   & Hi-GAL         &  1199.2   &  175.0   &   6.0     &  3.38  &   J/MNRAS/471/100/hcatalog \\
$350~\mu m$   & Hi-GAL         &  856.55   &   68.1   &   2.7     &  3.38  &   J/MNRAS/471/100/hcatalog \\
$350~\mu m$   & Hi-GAL         &  856.55   &   56.7   &   2.2     &  3.38  &   J/MNRAS/471/100/hcatalog \\
$500~\mu m$   & Hi-GAL         &  599.59   &   49.5   &   2.7     &  3.38  &   J/MNRAS/471/100/hcatalog \\
$500~\mu m$   & Hi-GAL         &  599.59   &   20.1   &   1.1     &  3.38  &   J/MNRAS/471/100/hcatalog \\
$870~\mu m$   & Hi-GAL         &  344.59   &   4.35   &   ""      &  3.38  &   J/MNRAS/471/100/hcatalog 
\enddata
\end{deluxetable*}

\end{document}